\documentclass{svjour3x}                     
\smartqed  
\usepackage{graphicx}
\newcommand{\gsim}{\,\raisebox{0.2em}{$>$}\!\!\!\!\!
\raisebox{-0.25em}{$\sim$}\,}

\newcommand{\gr}{$\gamma$-ray}
\newcommand{\grs}{$\gamma$-rays}
\newcommand{\rxj}{RX~J1713.7-3946}
\newcommand{\hess}{H.E.S.S.}
\newcommand{\byKB}{ (from K. Bernl\"ohr)}
\journalname{Experimental Astronomy}
\begin{document}

\title{Imaging Very High Energy Gamma-Ray Telescopes}


\titlerunning{Imaging VHE Gamma-Ray Telescopes} 

\author{Heinrich J. V\"olk         \and
        Konrad Bernl\"ohr
}

\authorrunning{H.J. V\"olk and K. Bernl\"ohr} 

\institute {H.J. V\"olk  
           \at Max-Planck-Institut f\"ur Kernphysik, P.O. Box 103980, 
           69029  Heidelberg, Germany\\
              \email{Heinrich.Voelk@mpi-hd.mpg.de}           
           \and
           K. Bernl\"ohr
           \at Max-Planck-Institut f\"ur Kernphysik, P.O. Box 103980, 
           69029  Heidelberg, Germany\\
           and \\
           Institut f\"ur Physik, Humboldt-Universit\"at zu Berlin, Germany \\
           \email{Konrad.Bernloehr@mpi-hd.mpg.de}
}

\date{Received: date / Accepted: date}

\maketitle

\begin{abstract}
  The technique of \gr\ astronomy at very high energies (VHE:$>100$~GeV) with
  ground-based imaging atmospheric Cherenkov telescopes is described, the
  \hess\ array in Namibia serving as example. Mainly a discussion of the
  physical principles of the atmospheric Cherenkov technique is given,
  emphasizing its rapid development during the last decade. The present status
  is illustrated by two examples: the spectral and morphological
  characterization in VHE \grs\ of a shell-type supernova remnant together with
  its theoretical interpretation, and the results of a survey of the
  Galactic Plane that shows a large variety of non-thermal sources. The final
  part is devoted to an overview of the ongoing and future instrumental
  developments.


\end{abstract}


\section{Introduction}
\label{intro}

At \gr\ energies $E_{\gamma}~ \gsim~ 30$~MeV electron-positron pair production
in the Coulomb field of an atomic nucleus dominates the interactions of photons
with normal matter. Cosmic \grs\ of such energies can therefore be most
effectively measured in either gas-filled or solid state satellite detectors
above the atmosphere registering the tracks of the resulting $e^{\pm}$-pairs, or
using the Earth's atmosphere itself as detector in observing secondary
Cherenkov radiation with ground-based telescopes. Satellite detectors work from
$\sim 30$~MeV up to some tens of GeV, where they become statistics-limited as a
result of the limited size of detectors that can be transported by rockets into
space. In the Very High Energy range (VHE; $E_{\gamma} > 100$~GeV) large
optical telescopes are used, often with diameters of 10~m or more.
\begin{figure*}
  \includegraphics[width=\textwidth]{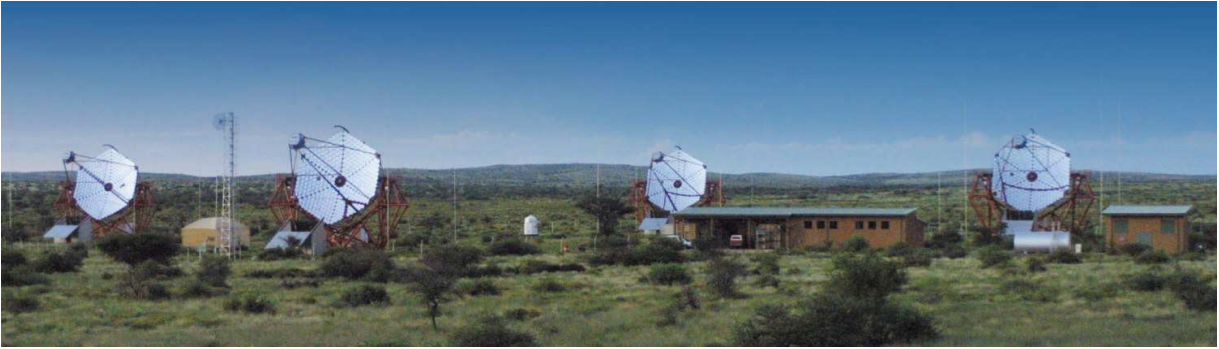}
\caption[The \hess\ telescope array]{The \hess\ telescope array \cite{hesspic} in the Khomas Highland of
  Namibia at an elevation of 1800 m a.s.l. The four 13~m telescopes stand at the
  corners of a square with a side length of 120 m.}
\label{fig:1}       
\end{figure*}

The history, development and results of satellite \gr\ detectors have been
described in this volume by K. Pinkau. Addressing the complementary ground-based
technique, we shall in the following summarize the physics of UV/Optical
imaging atmospheric Cherenkov telescopes which register the Cherenkov radiation
from the atmospheric secondary particles. Subsequently we shall discuss two
important astrophysical results that demonstrate the power of this observation
method. Given our personal involvement in the \hess\ (High Energy Stereoscopic
System) array in Namibia we shall in most instances use this experiment as
example (see Fig.~\ref{fig:1}), although we shall also emphasize the other
large arrays existing worldwide. The scientific results are astonishing, given
the slow development of the field over the three decades before the
mid-1990s. We conclude with a glimpse at the next generation European project
that envisages a further sensitivity increase by an order of magnitude.


\section{Air showers, secondary atmospheric Cherenkov emission, Imaging
  Atmospheric Cherenkov Telescopes}
\label{sec:2}

High-energy cosmic gamma rays produce $e^{\pm}$ pairs in the atmosphere over 
$\frac{9}{7}X_0$ (with the radiation length $X_0 \approx 36.5 \rm{g/cm}^2$
in air). This is followed by
Bremsstrahlung over the next radiation length which 
implies new \grs\ that generate new pairs, etc., until
the energy of the final generation of electrons becomes so small that their
fate is dominated by ionization losses which rapidly cool and thermalize
them. The result of these processes is called an electromagnetic Air Shower
that exists for about $10^{-4}$~seconds while traversing the atmosphere. The
multiplication of the number of particles and their eventual removal by
thermalization leads to a maximum number of shower particles at about 250 to
450 g/cm$^{2}$ for primary \grs\ of 20 GeV to 20 TeV, corresponding to an
atmospheric height of about 7 to 12 km above the ground.

The most realistic and complete description of the physical processes and the
corresponding results in the detector(s) is given by Monte Carlo (MC)
simulations. Except when explicitly noted we use the simulations performed by
one of us \cite{bernloehr2008} which use the CORSIKA code \cite{corsika}. It
incorporates all the physics of the atmospheric interactions and emission
processes and gives the possibility to statistically track their evolution in
the atmosphere, down into the detector.

\begin{figure*}
   \includegraphics[width=\textwidth]{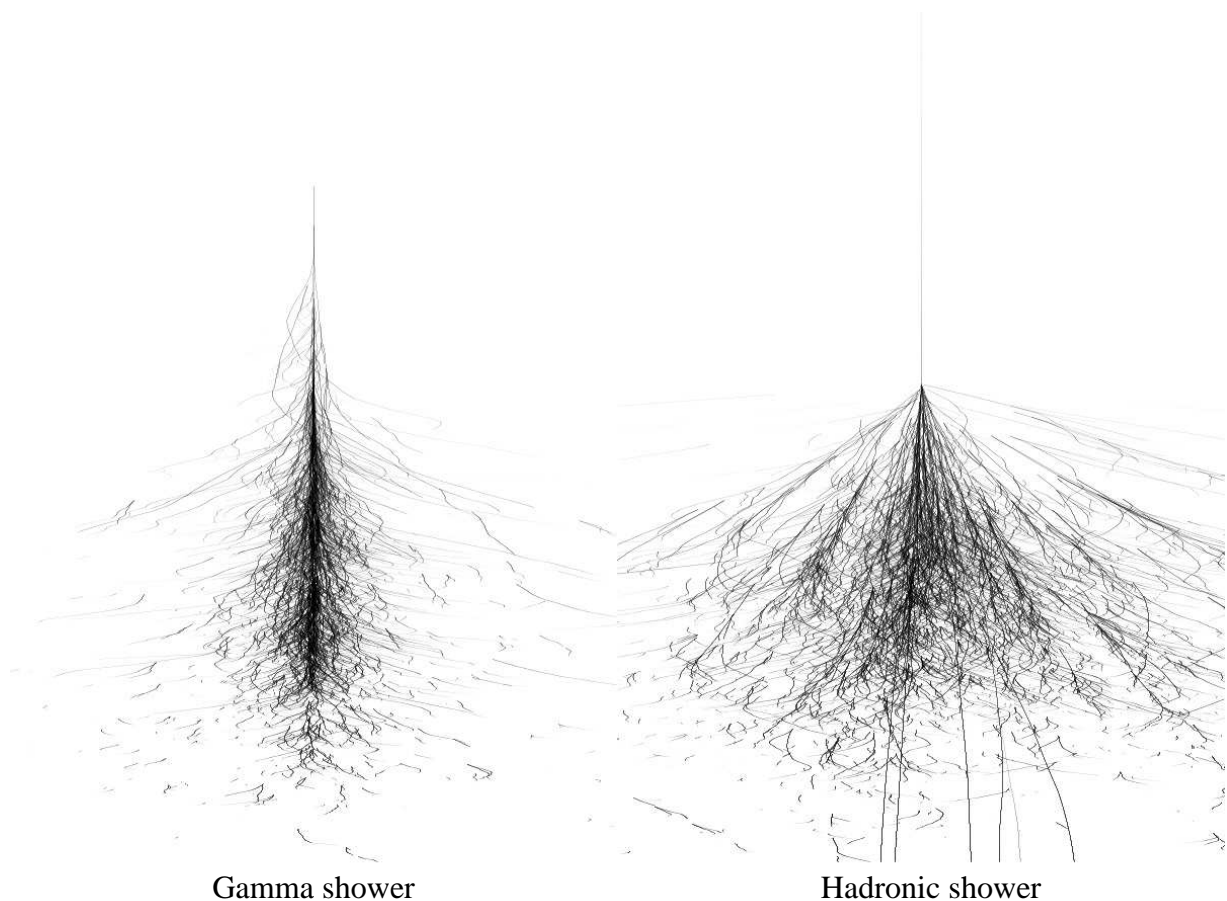} 
  \caption[Gamma showers and hadronic showers]{The different character of gamma showers and hadronic showers. The
    gamma shower is slender and to lowest approximation axially symmetric about
    the direction of the primary. The hadronic shower is irregular and may
    contain electromagnetic subshowers as a result of the large transverse
    momenta generated in hadronic interactions\byKB.}
\label{fig:2}       
\end{figure*}

Not only \grs\ penetrate into the atmosphere, but also charged energetic nuclei
(Cosmic Rays). They must be distinguished from the \grs. Energetic protons and
nuclei undergo hadronic interactions and produce dominantly neutral ($\pi^0$)
or charged ($\pi^{\pm}$) pions. Whereas the former decay into two gammas, the
latter ultimately produce electrons, positrons and two neutrinos via the $\pi~
\rightarrow \mu \rightarrow e$~decay. This leads to background air showers
which are of a mixed hadronic and electromagnetic nature. As a result
of the large transverse momentum transfer in hadronic interactions the hadronic
shower component is broad and irregular compared to the electromagnetic
component (see Fig.~\ref{fig:2}).

In general the flux of Cosmic Ray (CR) particles is much larger -- by a factor
of about $10^3$ -- than that of \grs. This implies a large background for \gr
 astronomy from the ground. It must be separated from the \gr\ signal, because
no anti-coincidence shield can be applied as in space detectors.

At primary \gr\ energies of about 100 GeV very few energetic photons or
electrons reach the ground. But the shower electrons from the original \gr\ are
still observable with optical telescopes through their Cherenkov radiation in
the optical range, because this atmospheric Cherenkov emission reaches the
ground without major absorption. Fig.~\ref{fig:4.1} below gives an impression
of the overall configuration. The disadvantages of this very promising
measurement technique are the weakness of the Cherenkov light and, to some
extent, its optical character. They require large light collection devices and
limit the observation time to clear and moonless nights. The observation
efficiency is typically about 10\%, depending critically on the astronomical
quality of the site.

\subsection{Cherenkov light pool}
\label{sec:2.1}

\begin{figure}
  \includegraphics[width=\textwidth]{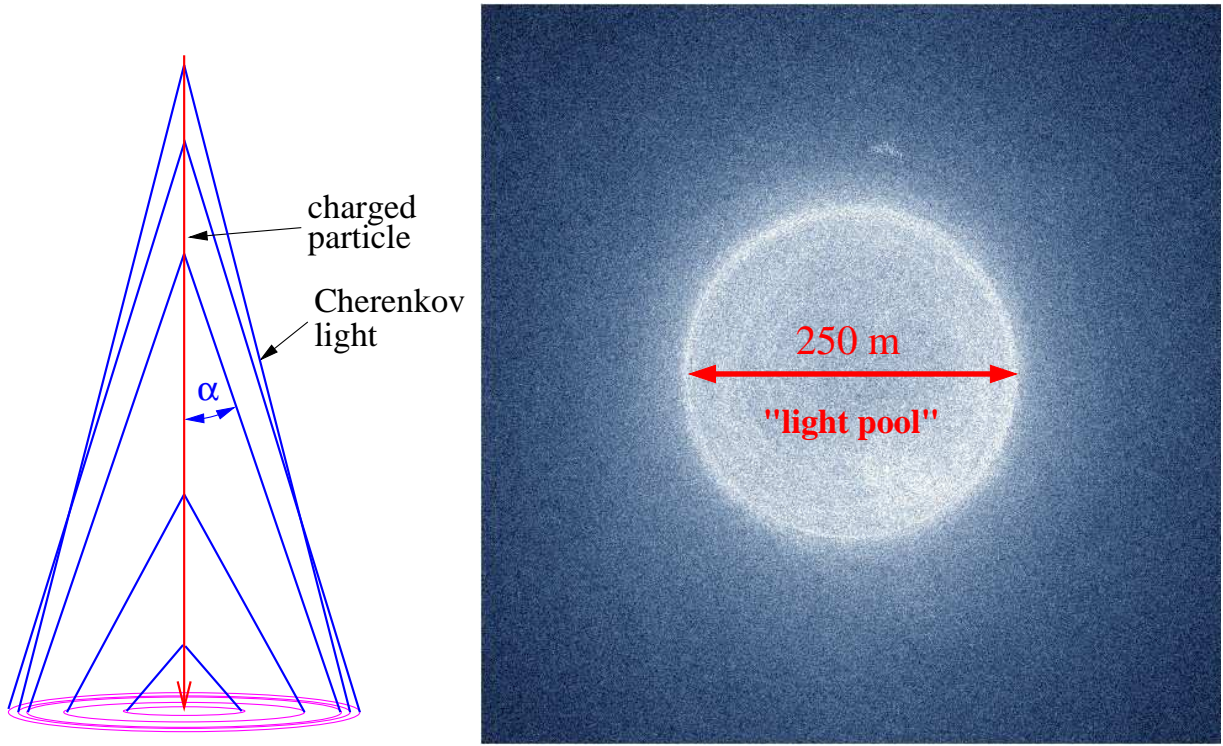}
  \caption[The ``light pool'']{Left: Atmospheric Cherenkov emission from a downward-moving single
    particle. Right: The ``light pool'' at an observation level at 1800 m
    above sea level from a \gr\ shower with a primary energy of 1 TeV\byKB.}
\label{fig:3}       
\end{figure}

The atmospheric Cherenkov light emission from a single particle is
characterized by a forward cone with an opening angle $\Theta\approx 1^{\circ}$
that increases downwards. For a particle moving vertically downwards, the
largest ring on the ground near sea level is from a height of 12 to 15~km (see
Fig.~\ref{fig:3}).

The ensemble of shower electrons from an energetic primary \gr\ produces a
rather uniformly illuminated ``light pool'' on the ground, centered on the
shower core, with a radius of about 125~m, if the multiple scattering of the
shower electrons is included (see Fig.~\ref{fig:3}).  To first
approximation it corresponds to the effective area of a telescope that images
the shower.  

\begin{figure}[ht]
\sidecaption
  \includegraphics[width=0.7\textwidth]{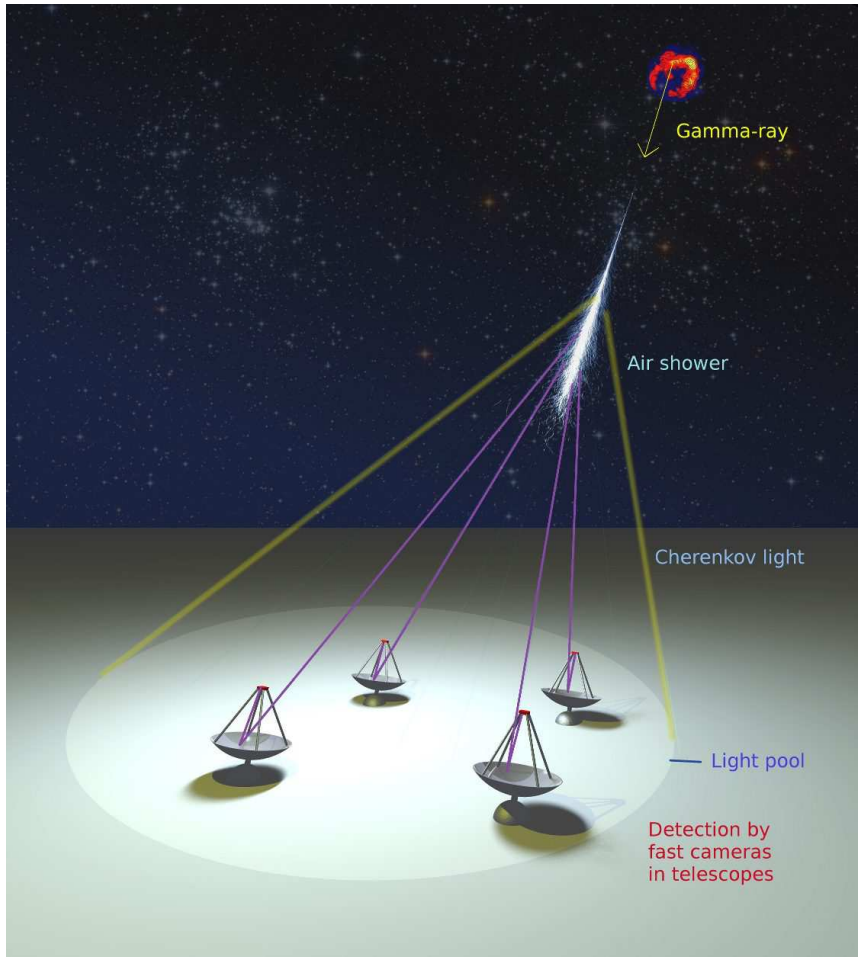}
  \caption[Schematic of the Cherenkov light pool]{Schematic of the Cherenkov light pool, originating from a primary
    \gr\ from within a cosmic-ray source (e.g. a supernova remnant) that
  illuminates an array of telescopes\byKB.}
\label{fig:4.1}       
\end{figure}

\begin{figure}[ht]
  \hbox to \textwidth{%
  \includegraphics[width=0.48\textwidth]{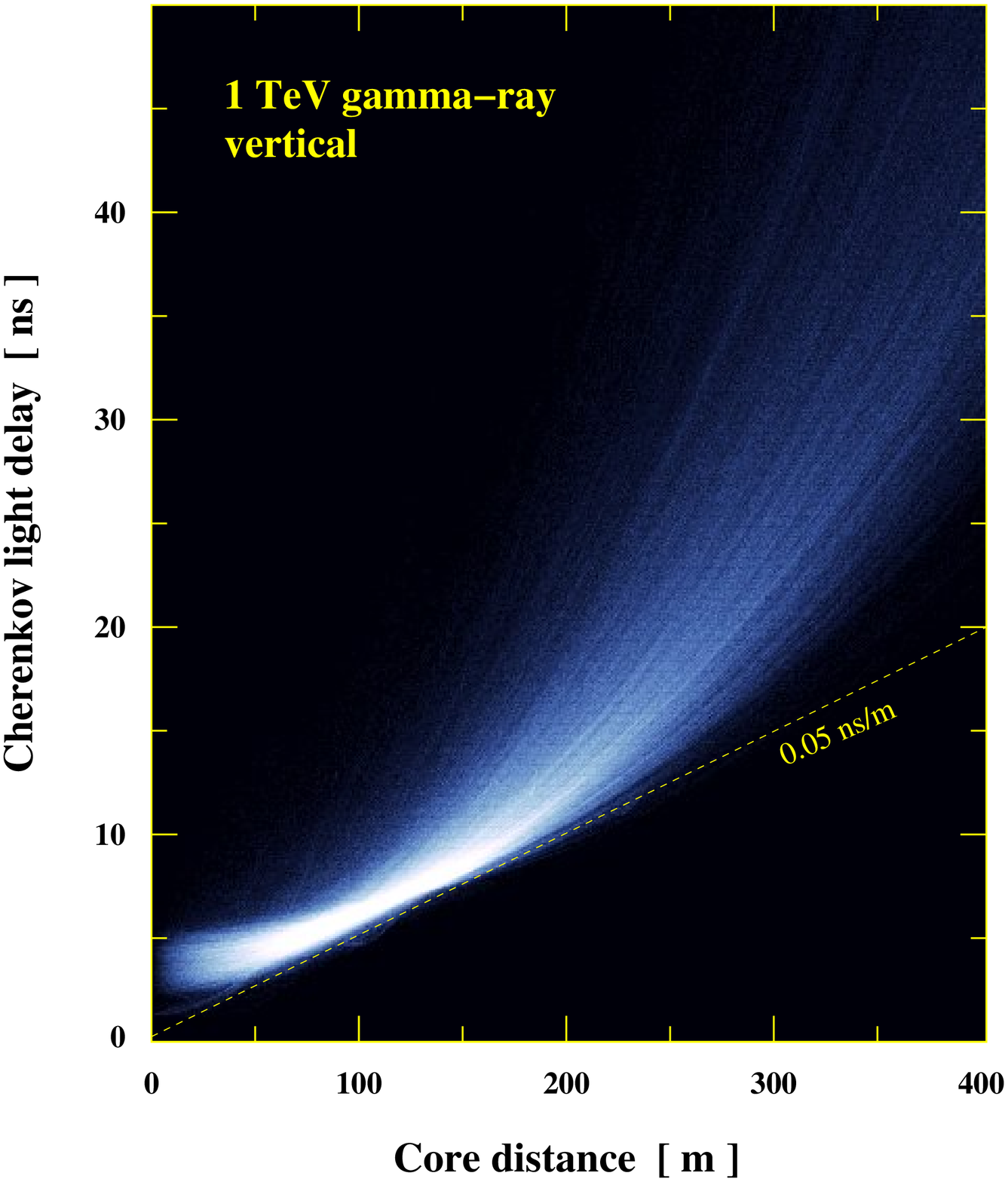}\hss
  \includegraphics[width=0.51\textwidth]{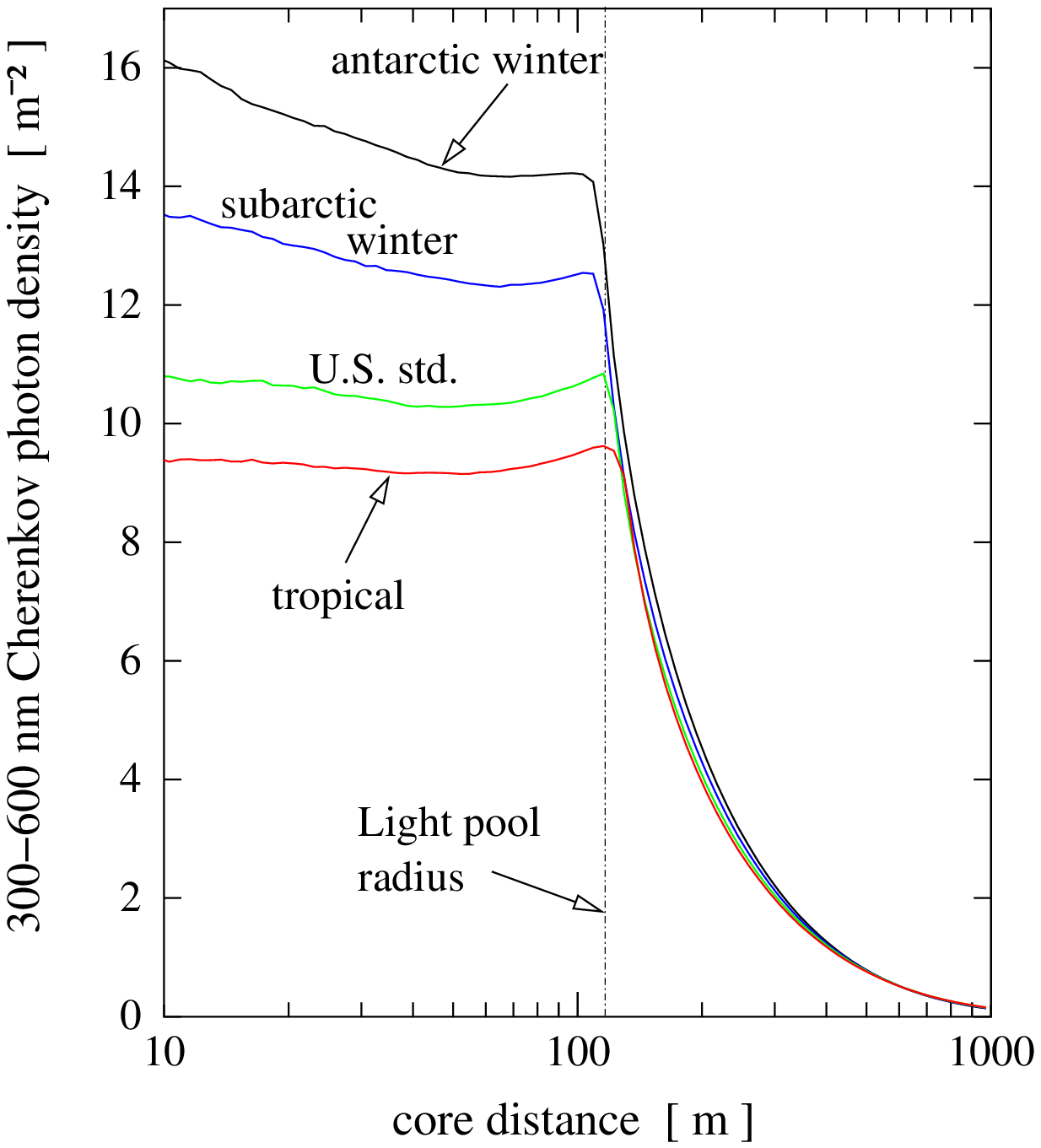}}
  \caption[Arrival time and lateral distribution]{Left: Distribution of arrival time and core distance of
     Cherenkov photons in a 1 TeV \gr-shower. Time is counted with respect
     to the time of arrival of the primary \gr\ on the ground 
     assuming no interactions. 
   Right: Lateral distribution per unit area of the optical Cherenkov emission
    from a shower with primary energy of 100 GeV, for various atmospheric
    profiles\byKB.}
\label{fig:4.2}       
\end{figure}

A schematic picture of a shower from a cosmic \gr\ source, illuminating an
array of telescopes on the ground, is given in Fig ~\ref{fig:4.1}. Since the
atmospheric index of refraction is very close to 1, the Cherenkov light almost
keeps pace with the radiating charged particles. Near the edge of the ``light
pool'' most light from a \gr\ shower arrives within 2~ns (see
Fig.~\ref{fig:4.2}). Thus a very short temporal window is possible, in order to
suppress the dominant night sky background. This suggests the use of fast
photomultiplier cameras for the telescopes.

\subsection{Atmospheric Cherenkov telescopes}
\label{sec:2.2}

The lateral distribution of the Cherenkov photon density depends somewhat on
the atmospheric profiles in its amplitude. However, the dependence on the
distance from the shower core - the extrapolation of the direction of the
primary \gr\ - is essentially independent of the atmospheric conditions. An
example is shown in Fig.~\ref{fig:4.2}. At a \gr\ energy of 100 GeV about 1000
Cherenkov photons are produced in a $100 \mathrm{m}^2$ telescope. With a
conversion efficiency of 10\% this results in $\sim 100$ photoelectrons in the
image. Given that the total number of Cherenkov photons is about proportional
to the primary \gr\ energy, this determines the threshold energy of the
telescope.

\begin{figure}
  \includegraphics[width=0.52\textwidth]{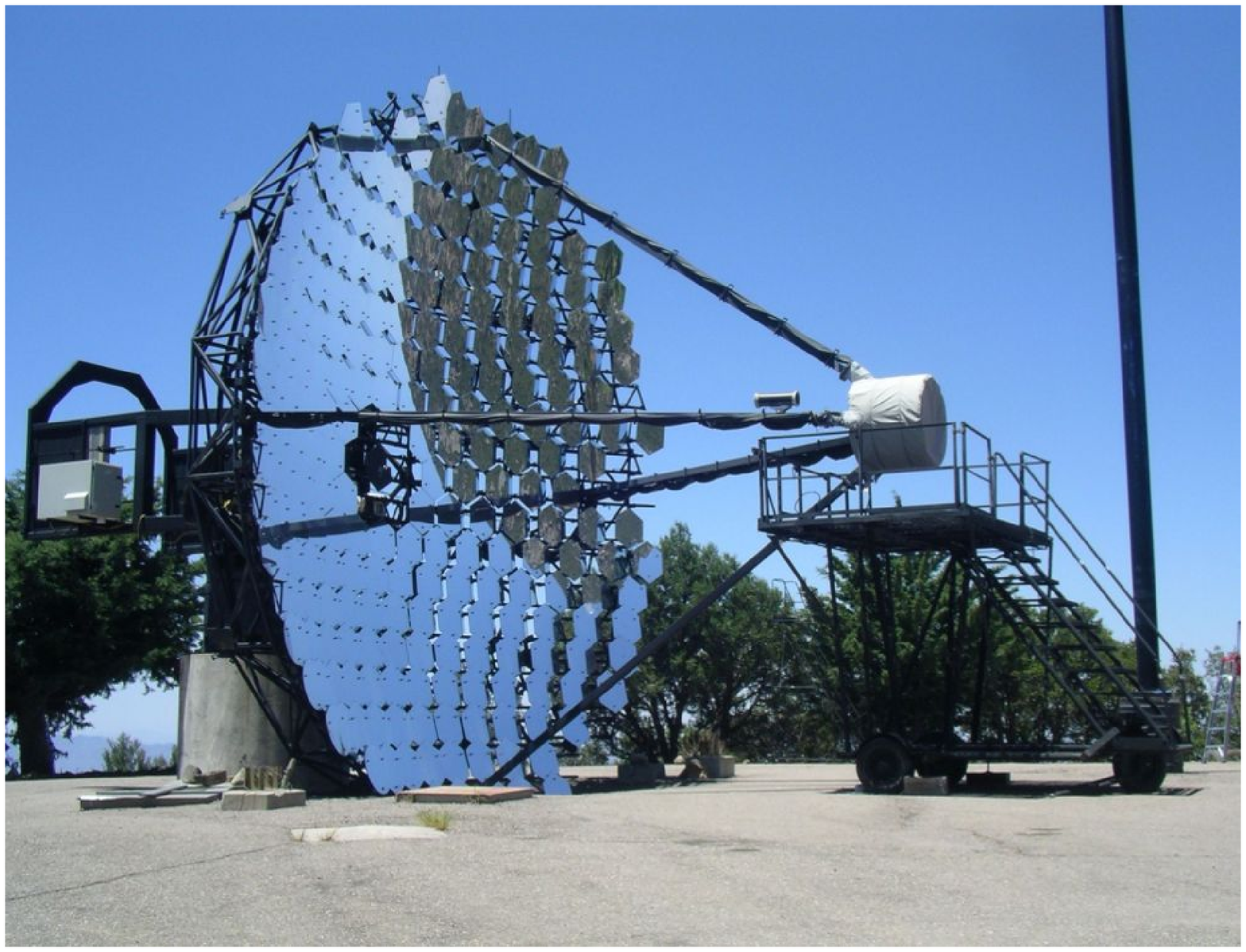}
  \includegraphics[width=0.48\textwidth]{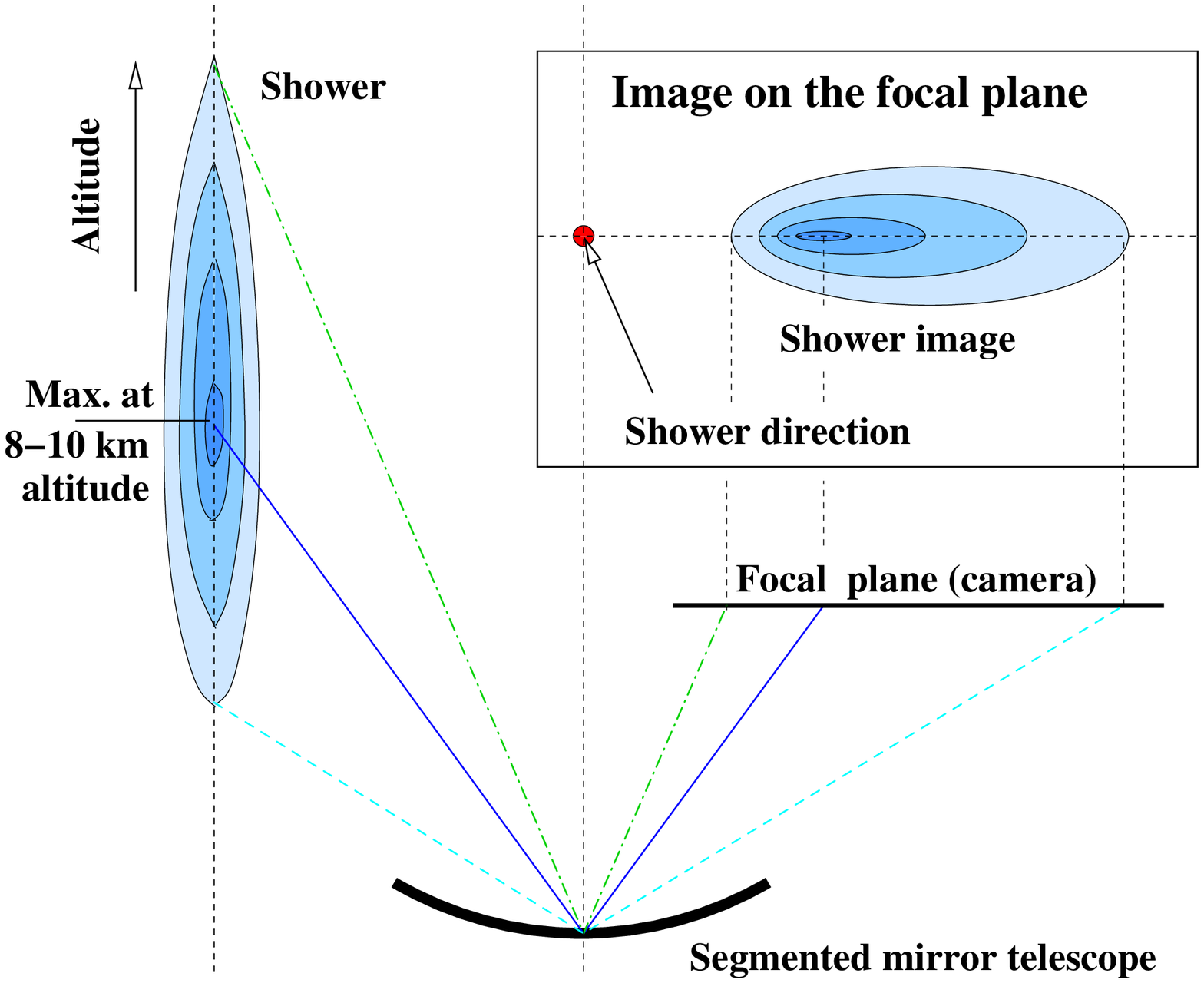}
  \caption[Whipple telescope]{Left: The 10~m Whipple telescope on Mt. Hopkins, 
    USA \cite{whipplepic}.  
    Right: Shower
    imaging by a telescope. Despite some asymmetry, the shower image in the
    camera has the shape of an ellipse. The shower direction is a point somewhere 
    on the extension of its major axis.
    For \gr\ primaries the image intensity gives
    the primary energy\byKB.}
\label{fig:5.2}       
\label{fig:6.1}       
\end{figure}

After successful initial experiments concerning CR air showers - what we call
today the CR background - in the 50ies in the UK \cite{galbraith53} and in the
USSR \cite{nesterova55}, the first dedicated \gr\ observations were attempted
in the Crimea by the Lebedev group from Moscow, following the suggestion by
\cite{zatsepin61}. At the time the results were basically negative
\cite{chudakov64}. Nevertheless they resulted in an upper limit to the VHE 
\gr\ flux from the Crab Nebula\footnote{A very interesting account of the early
  efforts in the USSR is given by A.S. Lidvansky in the talk ``Air Cherenkov
  Methods in Cosmic Rays: A Review and Some History '' presented at the
  centenary conference ``P.A. Cherenkov and Modern Physics'' (Moscow, 2004)~
  \cite{lidvansky06}.}$^,$\footnote{A condensed history of the field until 1994,
  very worthwhile reading, is to be found in \cite{weekes96}.}.

In the USA the growing interest in VHE \gr\ astronomy led to the construction
(in 1968) of a 10~m optical reflector with tessellated mirrors at the Whipple
Observatory in southern Arizona (see Fig.~\ref{fig:5.2})\footnote{Of the
  scientists originally involved (G.G. Fazio, J.E. Grindlay, G.H. Rieke,
  T.C. Weekes, and others) several have later also become leaders in fields
  like Infrared Astronomy and X-ray Astronomy.}. The imaging of the shower's
Cherenkov light is schematically shown in Fig.~\ref{fig:6.1}.

Despite intriguing indications the results remained controversial for about 20
years. With a multi-pixel photomultiplier camera in the focus since the
mid-80ies and the introduction of image analysis finally the Crab Nebula could
be detected with high ($9 \sigma$) significance after 60 hours of observation
time \cite{weekes89}.  Even after this breakthrough and a number of further
significant detections, the field continued to evolve rather slowly until the
mid-90s when a number of new telescopes started operation. They introduced
the stereoscopic technique and very fine camera pixelation (see below), leading
up to the present group of four major telescope systems worldwide.

\begin{figure*}
  \includegraphics[width=\textwidth]{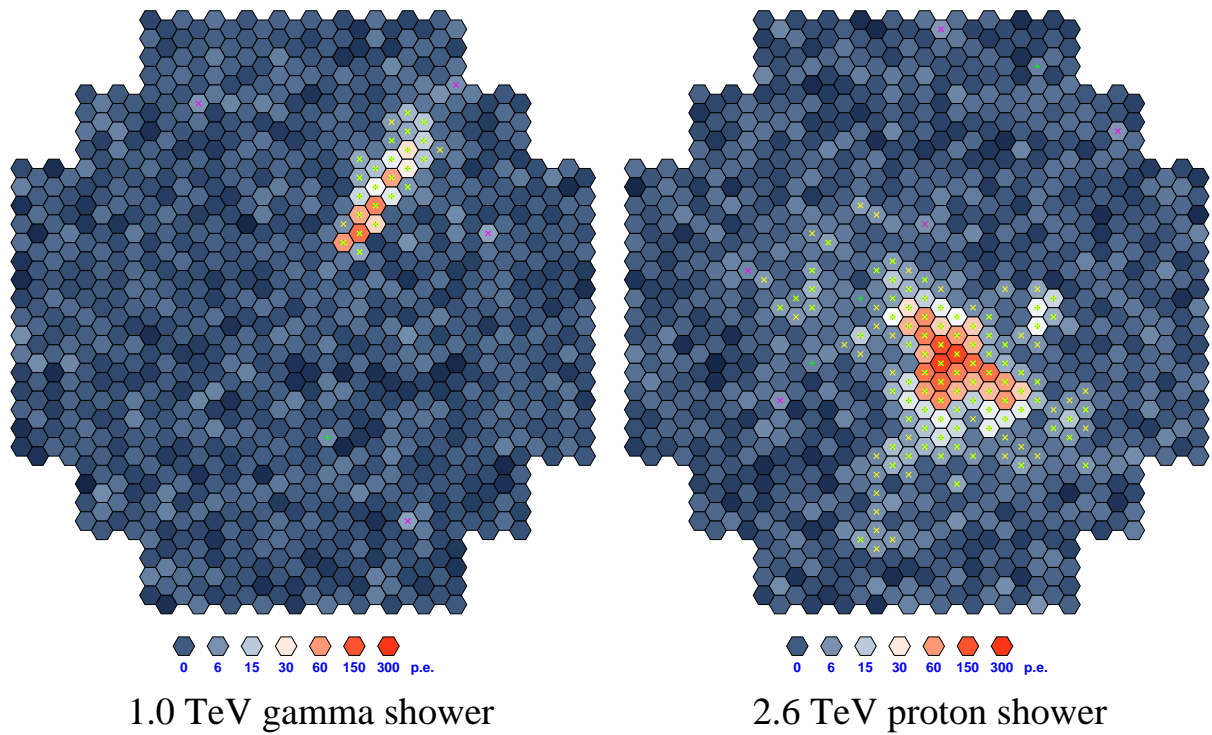}
  \caption[Images of gamma-induced and hadron-induced
      showers]{Difference between the images of gamma-induced and hadron-induced
      showers in the camera\byKB.}
\label{fig:7}       
\end{figure*}

The difficulties in the early \gr\ observations may be traced back to the
massive background of nuclear CR events. In addition, the large spatial
extension of air showers in the direction of the primary photon trajectory
leads to a very extended image $\sim 1^{\circ}$ in the camera plane (see
Fig.~\ref{fig:7}), making the size of the field of view (FoV) of the camera a
critically important parameter of the system. The lateral spread and
irregularity of hadronic showers increases this extension into the second
dimension and makes itself visible in the camera through an irregularly
structured image.

This difference between \gr showers and hadron showers can
be used to distinguish the nature of the generating particles through {\it
image analysis}, given a sufficiently dense array of photomultipliers in the
camera \cite{hillas85}.

\begin{figure*}
  \includegraphics[width=\textwidth]{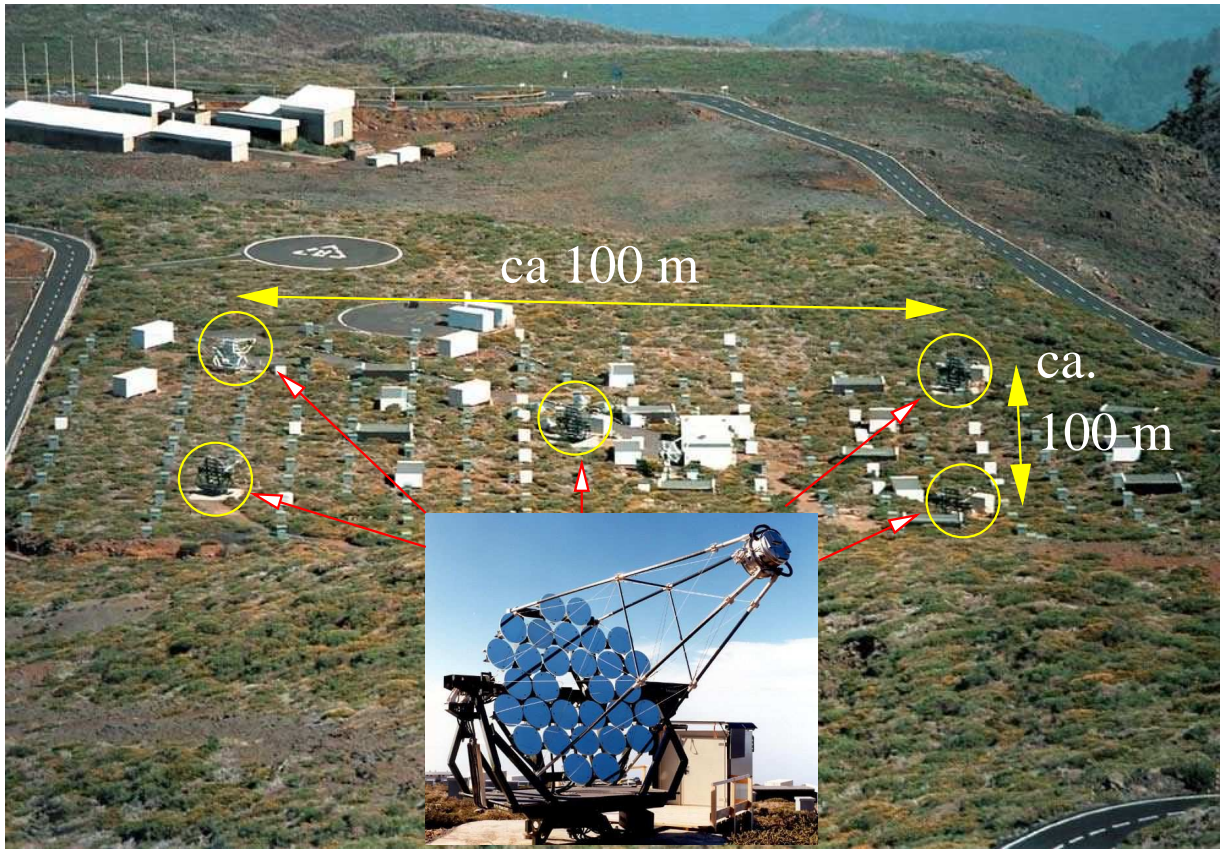}
  \caption[HEGRA]{The start of stereoscopy: HEGRA on La Palma (1995--2002)
    \cite{hegrapic}. The five
    3.5~m telescopes were situated in the center and at the 4 corners of a
    square of $~100$~m sidelength. The FoV was $\approx 5^{\circ}$.}
\label{fig:8}       
\end{figure*}

The aim of Hillas' algorithm was to reject the dominant events from CR nuclei
by suitable cuts on the images that are derived from Monte Carlo
simulations. The application of this method has proven to be very
successful. The remaining problem is that with a single telescope one obtains
only one projection of the Cherenkov light ``shower''. In addition, a single
telescope suffers from a second kind of background effect. It is produced by
penetrating muons (from CR interactions) that reach the ground at the telescope
mirror or its immediate neighborhood. These muon events are not easily
distinguishable from \grs. The solution for these drawbacks consists in the
stereoscopic technique. It was pioneered in the HEGRA telescope array on La
Palma \cite{daum97} that started full operation in the mid-90s (see
Fig.~\ref{fig:8}).

\subsection {Stereoscopic method}

\begin{figure*}
  \includegraphics[width=\textwidth]{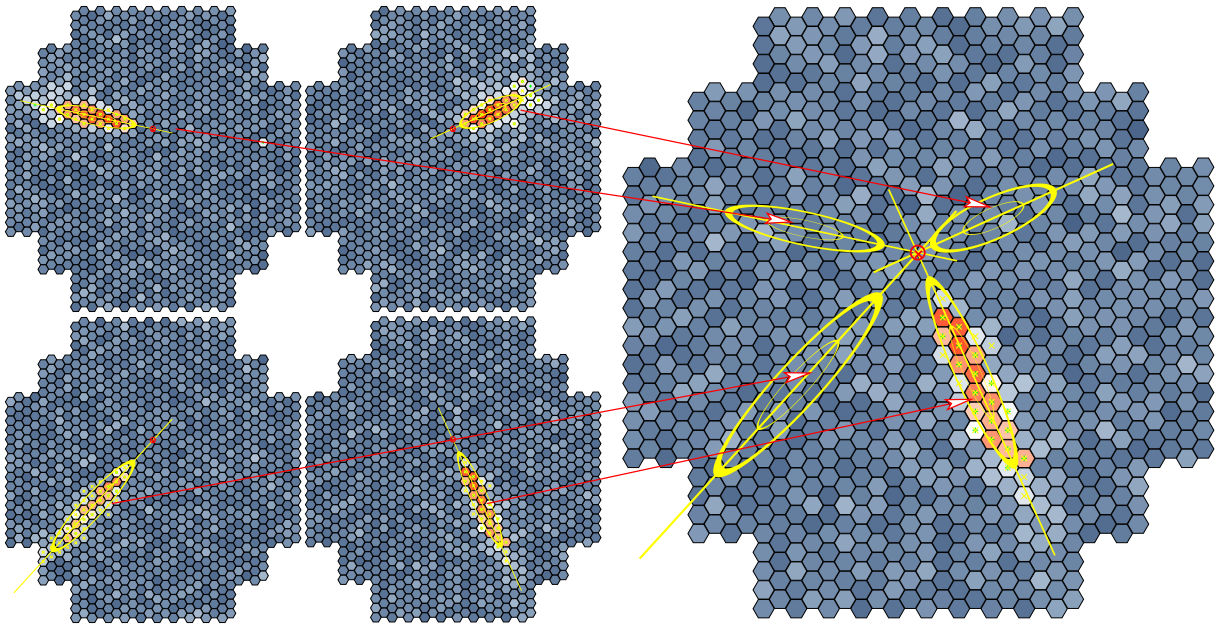}
  \caption[Stereo imaging]{Projection of several images from within the light pool of an event
   into one camera plane in a stereo system with four telescopes\byKB.}
  \label{fig:9}       
\end{figure*}

The use of several telescopes observing the same shower in coincidence allows a
unique determination of the shower direction by projecting the images in all
triggered telescope cameras into one camera (see Fig.~\ref{fig:9}). Then the
intersection point of the image major axes yields the shower
direction. Compared to a single telescope the angular resolution, the energy
resolution, the background rejection and the sensitivity are improved. In
addition this method allows the 3-dimensional reconstruction of the shower,
including the height of maximum particle number. The most advanced data
analysis methods use 3-dimensional modeling of the shower, no longer
confining themselves to the use of the 2-dimensional image parameters alone.

\begin{figure*}
  \includegraphics[width=0.48\textwidth]{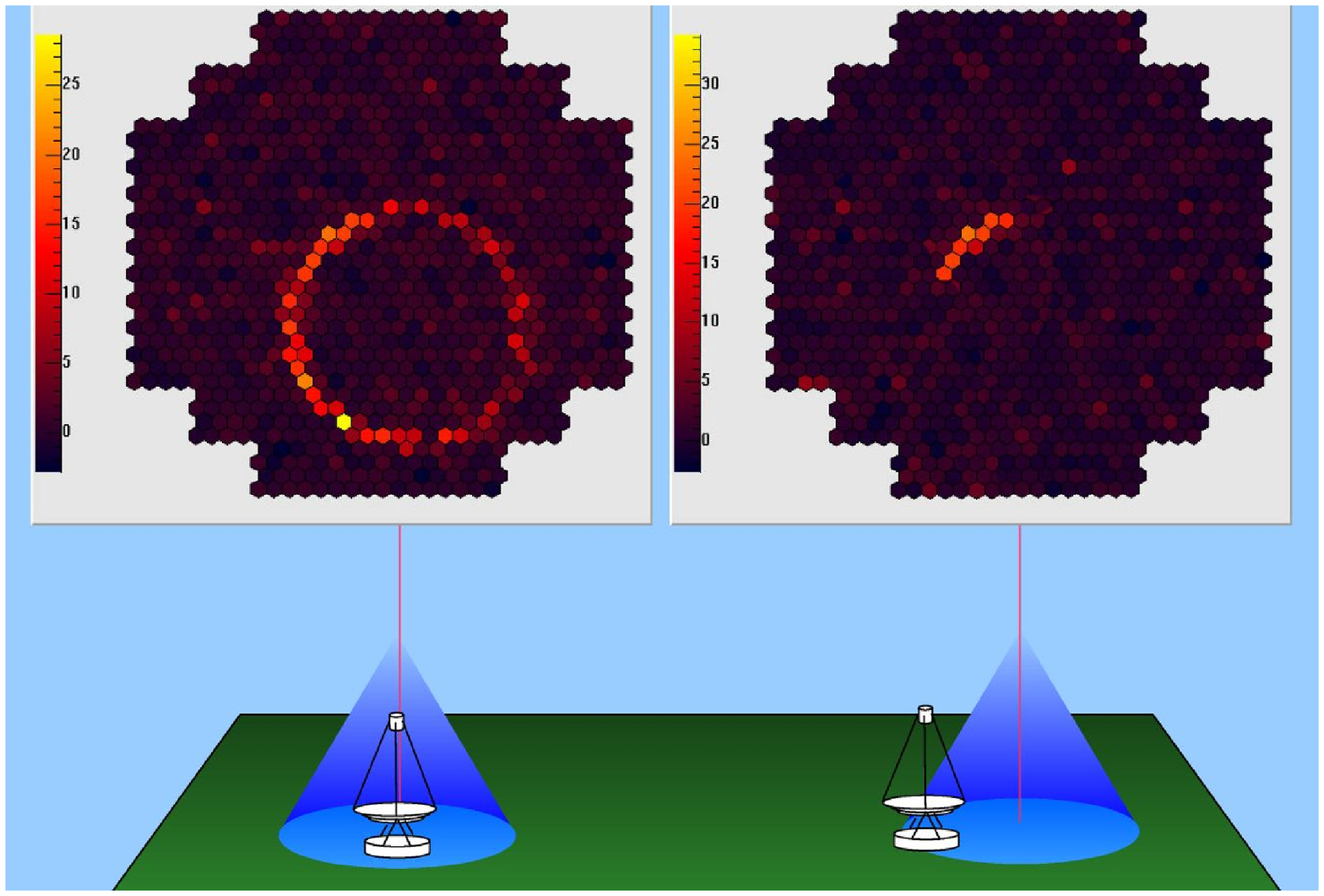}
  \includegraphics[width=0.52\textwidth]{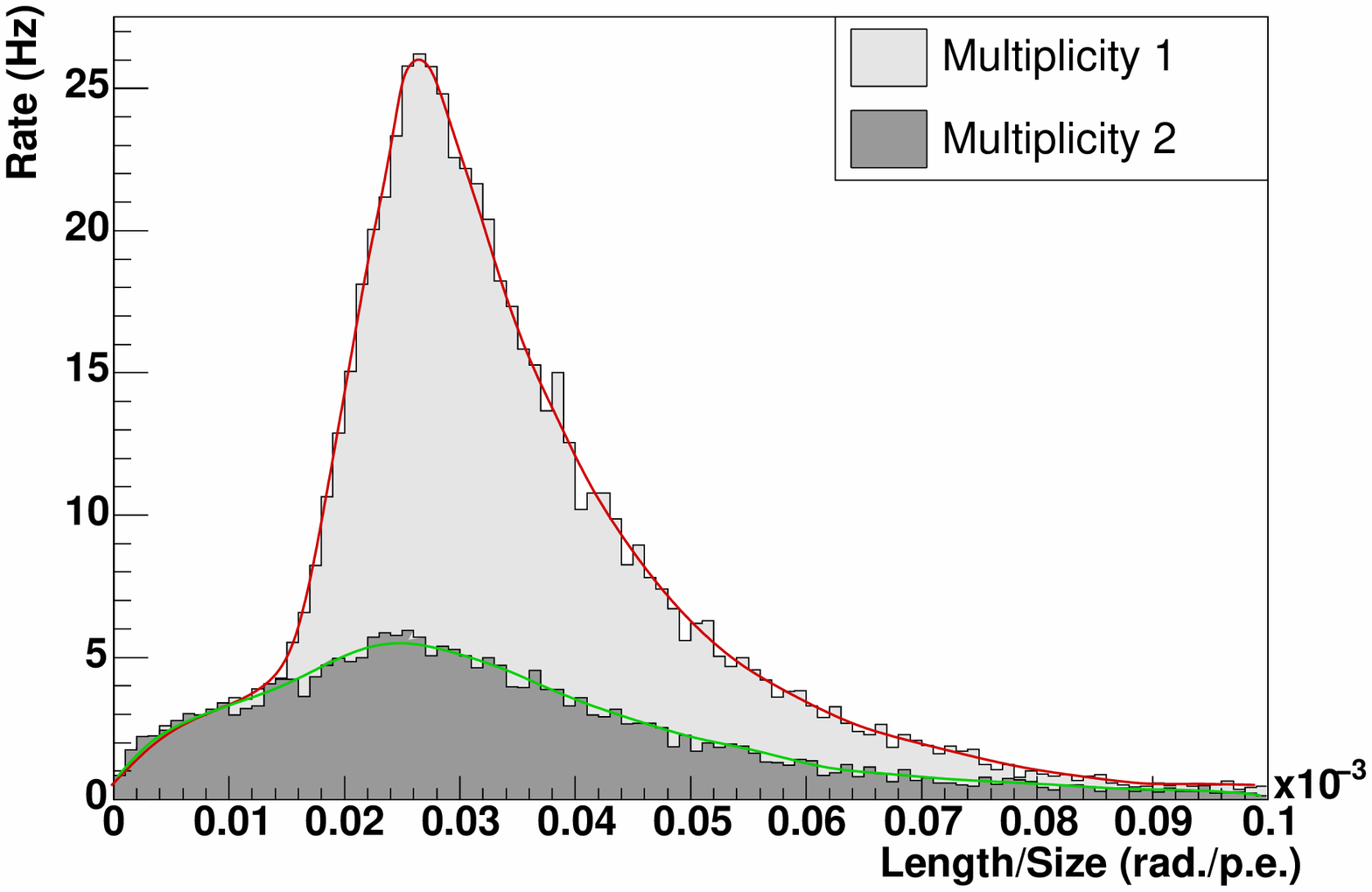}
  \caption[Local muons]{Local muons hitting telescopes in the center and close
    to the periphery (left, courtesy G.~Hermann and W.~Hofmann).  In peripheral
    encounters the resulting image is difficult if not impossible to
    distinguish from a low-energy \gr\ shower (right, based on H.E.S.S. data
    \cite{hesstrig}).  No such effect occurs in the other telescopes of a
    multiple system. The exclusion of muon events in stereoscopic systems
    reduces the muon background drastically.}
  \label{fig:10}       
\end{figure*}

The second major step forwards in the stereoscopic observation mode is the
suppression of the above-mentioned local muons with a stereo trigger: they
leave an image only in the telescope concerned, but not in the other telescopes
(unless the telescope light-gathering power is so enormous that even a single
charged particle can trigger an event from an inter-telescope distance, a case
which we will not discuss here). Therefore such events can be almost completely
eliminated (see Fig.~\ref{fig:10}). This is most important near the
energy threshold, where the shower images get weak and poorly defined. In other
words, only stereoscopic systems can reach the theoretical energy threshold
derived from the mirror size. Single telescopes are severely hampered by the
muon background and must operate significantly above their theoretical
threshold.

Despite the small mirror sizes of its components, the HEGRA array became the
most sensitive VHE instrument of its kind. It could thus prove the promise of
stereoscopy. Among others, several Blazar-type extragalactic sources, like the
active galactic nuclei Mrk 421 and Mrk 501, discovered shortly before by the
Whipple telescope, were confirmed or even measured with improved quality. The
\gr\ emission of the young Galactic Supernova Remnant Cassiopeia~A was
discovered in a series of deep observations over a period of 232~hours, and an
unidentified VHE \gr\ source in the Cygnus region was detected; its
astrophysical origin is subject of discussion to this day. In spite of these
important results the HEGRA array was not large enough to see more than the
``eight-thousanders'' of the VHE range. Its scan of the
Galactic scan revealed no new source and was in this sense not successful. The
other existing instruments of the time - the Japanese-Australian CANGAROO I
telescope at Woomera and the British Durham Mark 6 telescope in Narrabri, both
in the southern hemisphere in Australia, the French CAT telescope in the French
Pyrenees (championing very fine camera pixelation), and others, together with
the Whipple telescope - all shared this problem (for an overview of these
instruments see \cite{ong98}).

\begin{figure}
  \includegraphics[width=\textwidth]{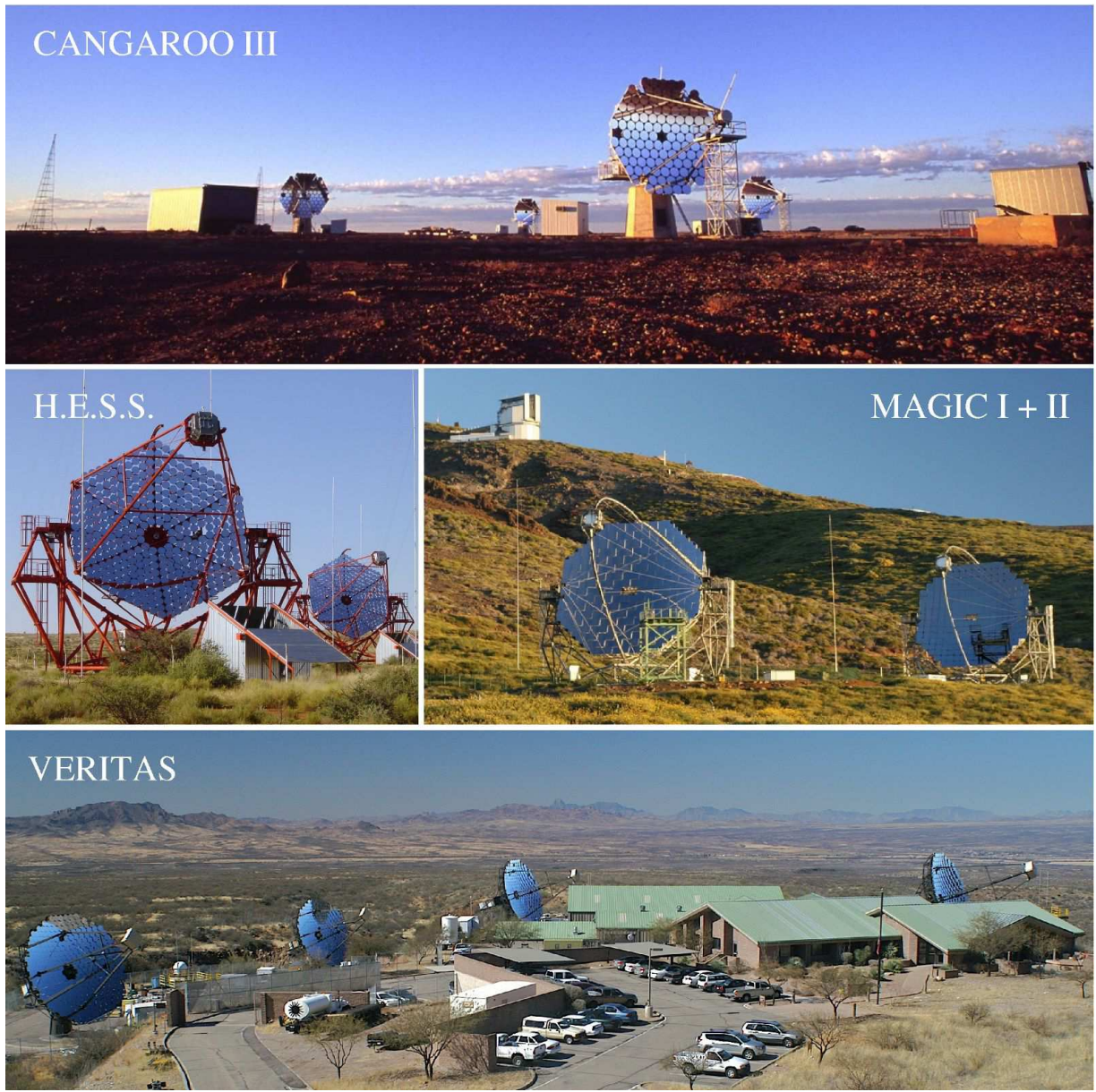}
  \caption[Third generation IACTs]{The third generation imaging Cherenkov telescopes: 
   the Japanese-Australian array CANGAROO III \cite{cangaroopic}
   the two European initiatives H.E.S.S. \cite{hesspic}
   and MAGIC \cite{magicpic}, 
   and the US-observatory VERITAS \cite{veritaspic}.}
  \label{fig:11}       
\end{figure}

\subsection {Third generation instruments; \hess}

As a consequence, a third generation of imaging Cherenkov telescopes was
envisioned and eventually realized. These were the VERITAS array
\cite{veritas03}, planned for the Mt. Hopkins area again, CANGAROO III
\cite{mori05} near Woomera, \hess\ in Namibia \cite{benbow05} - also in the
southern Hemisphere - and MAGIC \cite{lorenz05} on La Palma at the HEGRA site.
They all were to have high-definition cameras and, with the exception of MAGIC
that was initially designed and operated until recently as a large single
telescope, they were stereoscopic arrays, following HEGRA (see
Fig.~\ref{fig:11}).

As an example let us consider H.E.S.S. (High Energy Stereoscopic
System)\footnote{The conceptual basis for stereoscopic systems of $\sim 10$~m
  telescopes is given in \cite{AHKV97a,AHKV97b}.}. The experiment's name is at
the same time a tribute to Victor F.\ Hess, the Austrian physicist who
discovered the Cosmic Rays in 1912 in a series of carefully instrumented
high-altitude balloon flights. He received the Nobel prize for this discovery
in 1938.

The H.E.S.S.\ Collaboration involves a total of about 150 scientists from
various European countries, Armenia, South Africa and Namibia. While the
astrophysicists play a decisive role, the majority are physicists that adapt
techniques from accelerator physics to this young field of astronomy. The
pattern is reminiscent to that visible in the development of radio and X-ray
astronomy some decades earlier.

\begin{figure}\sidecaption
  \includegraphics[width=0.7\textwidth]{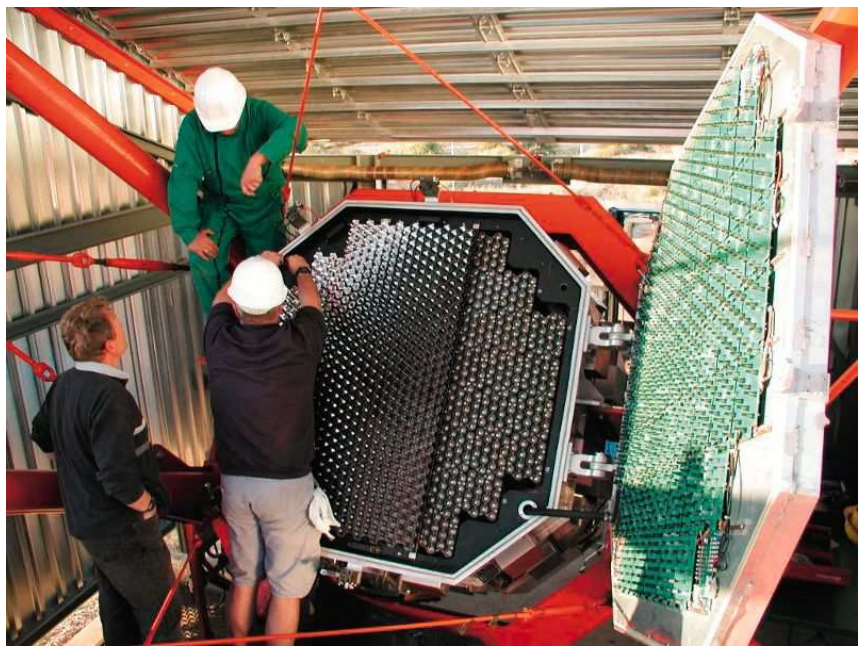}
  \caption[One of the \hess\ cameras]{One of the \hess\ cameras in its shelter \cite{hesspic}. It has 960
    photomultiplier pixels of size $0.16^{\circ}$, resulting in a
    $5^{\circ}$~field of view. The electronics with 1 GHz analog sampling rate
    is entirely contained in the camera body.}
  \label{fig:12}       
\end{figure}

The 4-telescope array (see also Fig.~\ref{fig:1}) is designed for maximum
mechanical stability despite full steering and automatic remote alignment
capability of the 382-piece mirror system that implies an area of 
107~m$^2$ per telescope; the individual glass mirrors have a diameter of 
60~cm. The camera in the primary focus weighs almost 1 ton (see
Fig.~\ref{fig:12})

As mentioned before, the large field of view of $5^{\circ}$ of the telescope
cameras is a key feature of \hess\.~Even though quite expensive in cost,
operation and maintenance, this requirement resulted from the realization that
most nearby Galactic VHE sources should be extended by $\sim 1^{\circ}$ degree
or more, e.g. \cite{drury94}. This should be also the case for the VHE emission
from nearby rich clusters of galaxies, one of the key targets of extragalactic
\gr\ astronomy \cite{voelk96}.

Only a large field of view permits efficient observations of such extended
objects. In addition it makes it possible to perform an efficient sky survey, especially
in the Galactic Plane. Beyond that, the collection area of the telescopes
increases with increasing \gr\ energy, since the Cherenkov light pool extends
beyond 125 m for very energetic showers that produce a large amount of
Cherenkov light. For the \hess\ telescopes this implies a considerable increase
of the collection area with energy up to \gr\ energies of about 100~TeV.

\begin{figure*}[b!]
\hbox to \textwidth{%
  \includegraphics[width=0.48\textwidth]{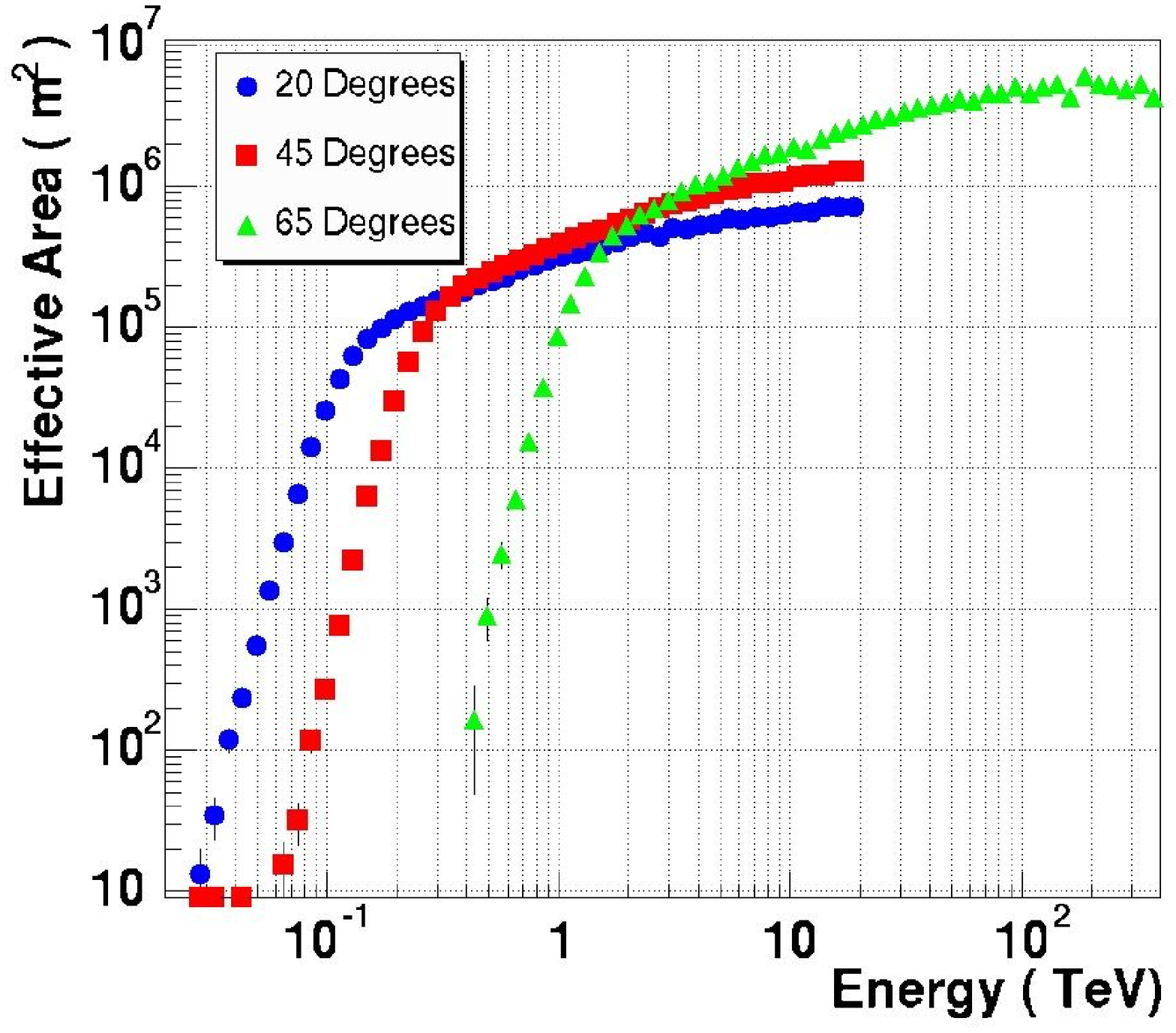}\hfill
  \includegraphics[width=0.48\textwidth]{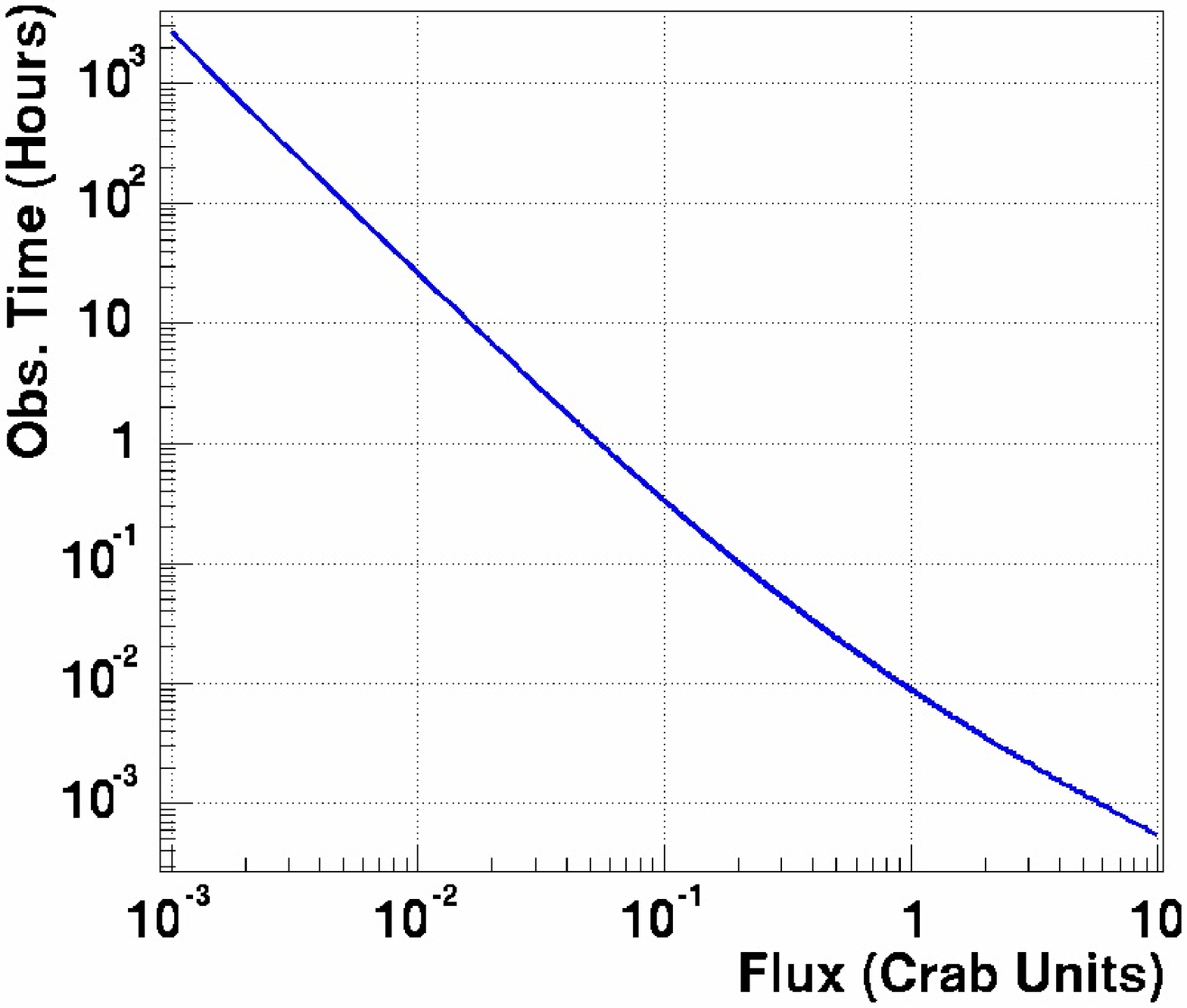}}
  \caption[Performance of \hess]{Performance of the \hess\ 4-telescope system. The effective area
    as a function of \gr\ energy is shown (left). The right panel gives the
    observation time required for a {5 $\sigma$} detection as function of the
    \gr\ flux in units of the flux from the Crab Nebula (\cite{benbow05,hesscrab}).}
  \label{fig:13}       
\end{figure*}

The resulting performance of the \hess\ array is summarized in
Fig.~\ref{fig:13}. After suitable data cuts on the expected direction of the
source and on the shape of the images, derived from Monte Carlo simulations
\cite{hesscrab}, the analysis accepts more than 50\% of the gamma events and
less than 0.1\% of the events from CR nuclei; the angular resolution is better
than $0.1^{\circ}$~per event and the energy resolution is $10 - 15$\%~per
event. This compares favorably with satellite detectors. The energy threshold
(at zenith) is 100 GeV. The collection area increases with \gr\ energy,
reaching $> 1 \mathrm{km}^2$ for observations at large zenith angles.  The
sensitivity in the overlapping energy range is a factor of $\approx 10$ better
than that of HEGRA and exceeds that of the single-telescope MAGIC by a factor
$\sim 3$. This implies an observation time of 1 hr for a $5 \sigma$ detection
of an energy flux of $10^{-11}$~and $10^{-12}$~erg cm$^{-2}$s$^{-1}$~at \gr\
energies of 100 GeV and 1 TeV, respectively. In other words, an instrument like
\hess\ can detect the (point-like) Crab Nebula in TeV \grs\ in about 30~s,
compared with the $\sim 18$ hr required for the equivalent original detection
15 years earlier\footnote{The Crab Nebula is generally treated as standard
  candle in VHE \gr\ astronomy.}.


\section{Two major scientific results}

In this section we shall illustrate the technique of ground-based \gr\ astronomy
with two examples from the \hess\ results. They share the ubiquitous
feature of spatial extension, characteristic of so many of the non-thermal
sources in the Universe. The first example is a Supernova Remnant (SNR), thus
possibly belonging to the long-sought class of sources of the Galactic CRs. The
other example concerns the results of the Galactic Plane scan in the VHE range
that has been performed since 2004.

\begin{figure}
  \includegraphics[width=\textwidth]{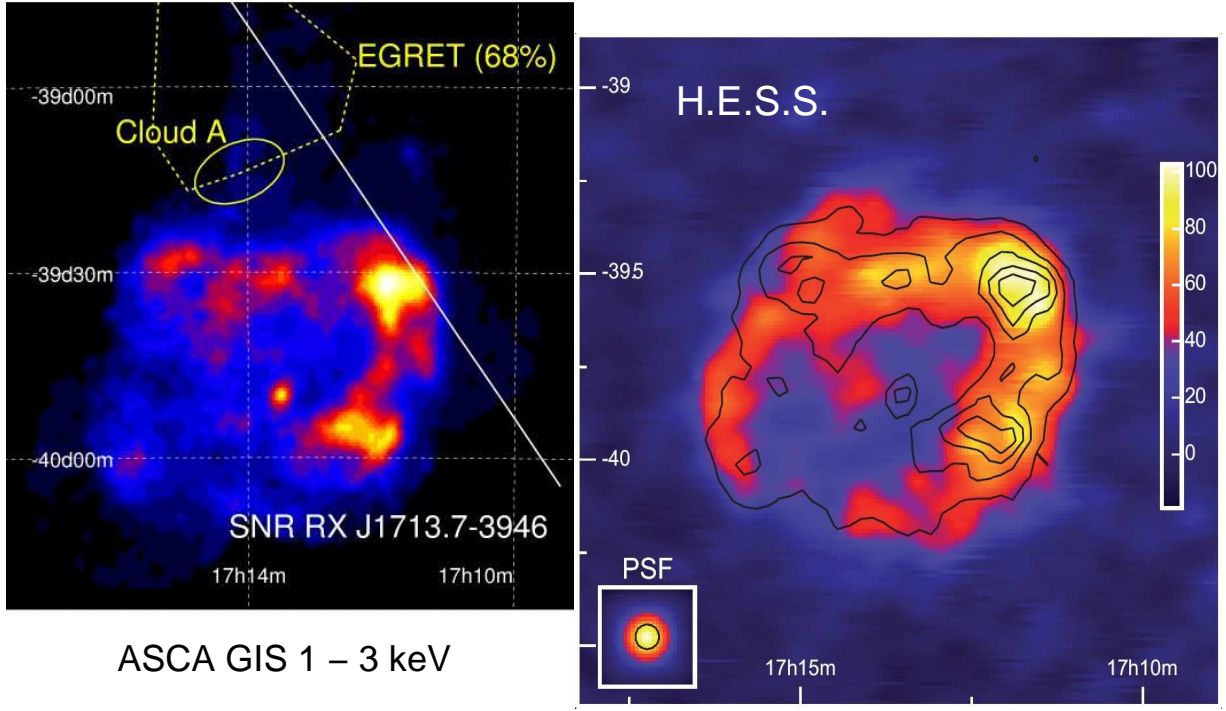}
  \caption[\rxj]{Left: Image of the Supernova Remnant \rxj, as observed with the
    ASCA satellite in the 1-3 keV band \cite{uchiyama02}.  Right: \hess\ image of \rxj\ at
    energies above 190 GeV (in false colors), with the ASCA contours
    superposed \cite{aha07}.}
  \label{fig:14.1}       
  \label{fig:14.2}       
\end{figure}

\subsection{The supernova remnant RX J1713.7-3946}

This SNR, located in the constellation Scorpius in the Galactic plane, was
originally discovered with the X-ray satellite ROSAT \cite{pfeffermann96}. In
X-rays \rxj\ has a diameter of $1^{\circ}$, twice the size of the full moon
(see Fig.~\ref{fig:14.1}). The ASCA (1-3 keV) image shows an overall shell
structure (\cite{uchiyama02}). Also in subsequent observations with Chandra and
XMM the X-ray emission was ``entirely'' non-thermal: no thermal emission has yet
been identified. The existence of a central X-ray point source, which is
probably a neutron star, suggests the core collapse of a massive star. \rxj\ was
also detected with \hess\ \cite{aha04,aha06} and there exists a close spatial
correlation of the X-ray and the VHE \gr\ emission (see Fig.~\ref{fig:14.2}).

\begin{figure}
\hbox to \textwidth{%
  \includegraphics[width=0.43\textwidth]{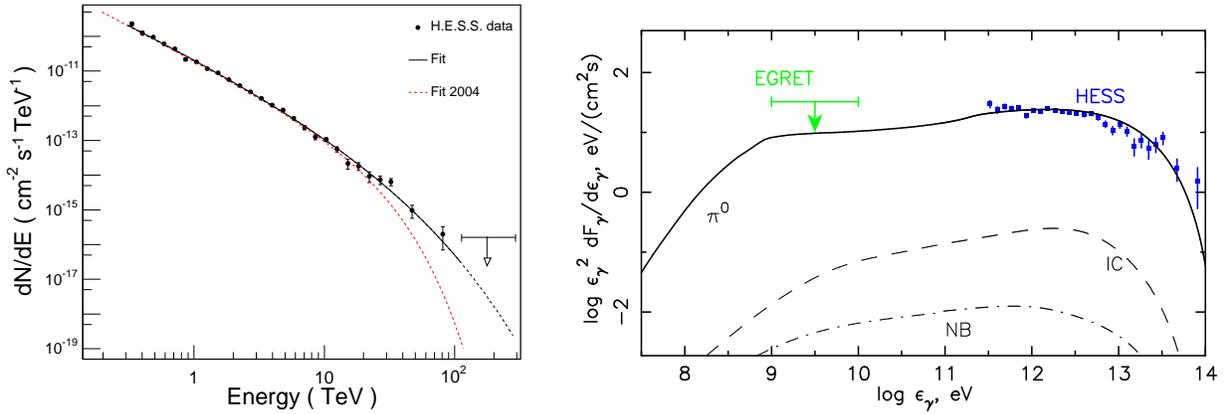}\hfill
  \includegraphics[width=0.53\textwidth]{0444fig1.eps}}
  \caption[Spectrum of \rxj]{Left: \hess\ energy spectrum of \rxj\ \cite{aha07}.
   Right: Theoretical spectral energy distributions of \rxj\ in the VHE range,
   assuming a magnetic field strength $B_{\mathrm{eff}} \approx 130 \mu$G
   inside the remnant \cite{bv08}.}
  \label{fig:15.1}       
  \label{fig:15.2}       
\end{figure}

The \gr\ spectrum is very hard - harder than a spectrum containing equal energy
per decade in the observed range - and extends up to \gr\ energies of about 100
TeV \cite{aha07} (see Fig.~\ref{fig:15.1}). This means that the generating
charged particles must have energies in excess of about 300 TeV and, in the case of
nuclear particles, even of about 600 TeV. A reasonable phenomenological fit to the
differential \gr\ energy spectrum is $dN/dE = N_0 E^{-\Gamma}
\exp{(E/E_\mathrm{c})^{0.5}}$, with $\Gamma = 1.8 \pm 0.04$ and $E_\mathrm{c} =
3.7$~TeV. Extrapolating this spectrum back to 1 GeV, using arguments from
diffusive shock acceleration theory (see below), yields a total energy in
energetic particles of about $10^{50}/{\langle n \rangle}$~erg. For an average
gas density of $\langle n \rangle \approx 1 \mathrm{cm}^{-3}$~ this
corresponds to roughly 10 percent of the expected total hydrodynamic explosion
energy. Also spectral imaging and an energy-resolved morphological
characterization was achieved.
 
\begin{figure*}
\begin{center}
  \includegraphics[width=0.85\textwidth]{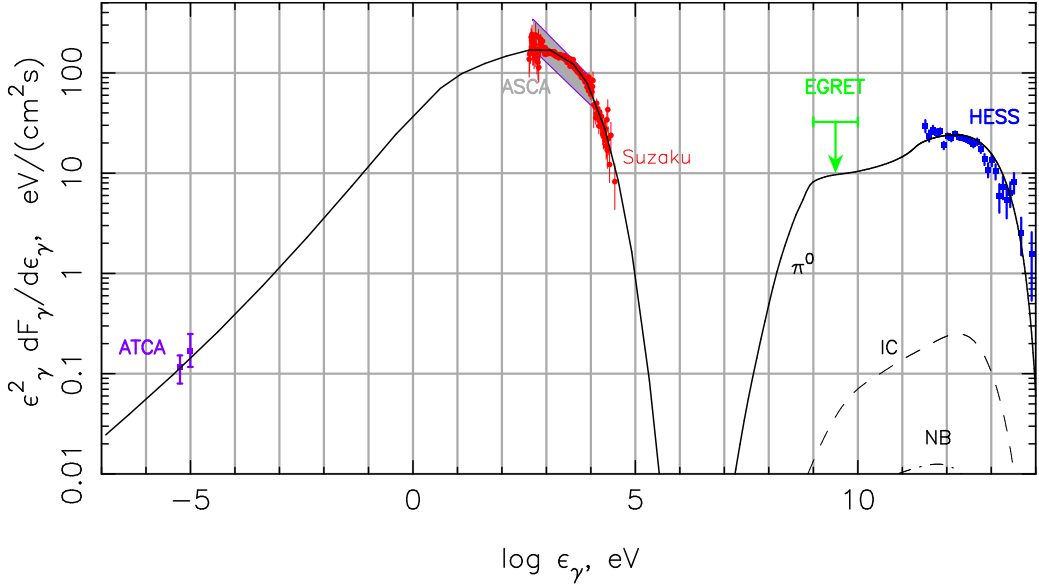}
\end{center}
  \caption[Energy distribution of \rxj]{Overall and spatially-integrated non-thermal spectral energy
   distribution of \rxj. The lower-energy synchrotron observations in the radio
   (ATCA) and X-ray (ASCA and Suzaku) ranges is compared with the theoretical
   model \cite{bv08}. The \gr\ spectrum at higher energies is the same as in
   Fig.~\ref{fig:15.2}. \\ ~ \\[1.5ex] ~}
  \label{fig:16}       
\end{figure*}

Given the possible fundamental significance of such an empirical finding we
show a comparison of the spatially-integrated spectral energy distribution
(SED) from \hess\ with a theoretical model spectrum \cite{bv06,bv08} that
assumes a magnetic field strength $B_{\mathrm{eff}} \approx 130 \mu$G inside
the remnant and is able to fit the {\it amplitude } of the VHE spectrum (see
Fig.~\ref{fig:15.2}) for a typical injection rate of supra-thermal nuclear
particles into the acceleration process. The inverse Compton (IC) emission and
Bremsstrahlung (NB) of the simultaneously accelerated CR electrons turn out to
be negligible in comparison with the \gr\ emission generated by the
$\pi^0$-production and subsequent decay into two \grs\ from inelastic
collisions of the accelerated nuclear particles with thermal gas nuclei in the
interior of the remnant. The upper limit in the GeV range, derived for this
region of the sky from the EGRET detector, is shown as well. The IC and NB
amplitudes in Fig.~\ref{fig:15.2} are derived from the observed synchrotron SED
in the radio and X-ray ranges below 100 keV (see Fig.~\ref{fig:16}), consistent
with the magnetic field strength above.

That 10 or more percent of the entire mechanical explosion energy of the
Supernova explosion should be transformed into non-thermal energy of
ultra-relativistic particles is a rather outrageous proposition from the
theoretical side. Yet nature might indeed comply with it, as the \hess\ result
shows. The predicted hadronic dominance of the \gr\ SED is expected to be
confirmed by the recently launched Fermi Gamma-ray Space Telescope with an SED
that should be only a factor $\approx 2.5$~lower at 1 GeV than that measured at
1 TeV from Fig.~\ref{fig:15.2}.

\begin{figure*}[htb]
  \includegraphics[width=\textwidth]{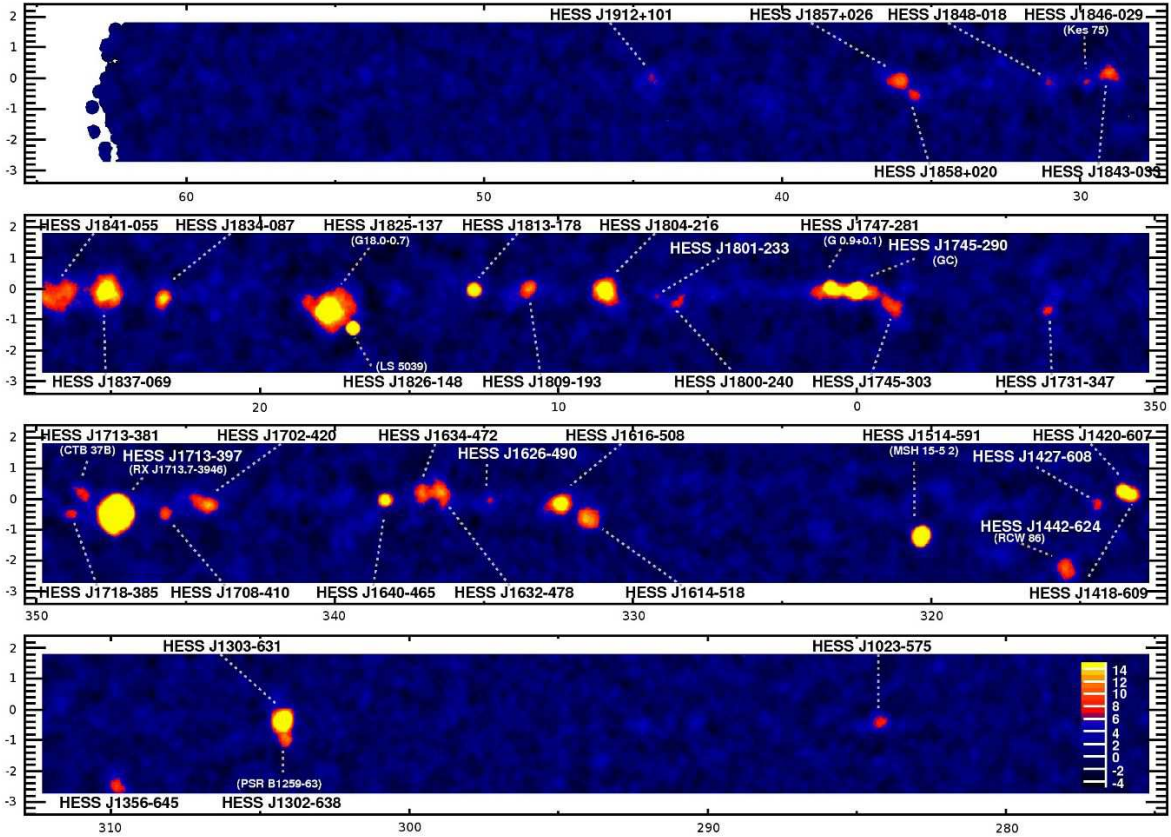}
  \caption[Galactic Plane Survey by \hess]{Significance map of the Galactic
    Plane Survey by \hess, presented in Galactic coordinates: $280^{\circ} < l
    < 60^{\circ}$~ and $-3^{\circ} < b < 3^{\circ}$ \cite{galscan07}. Most of
    the sources are spatially extended beyond the $0.1^{\circ}$ resolution of
    the array. The Galactic Center (GC) region and the SNR \rxj\ can be seen in
    the right half of the 2nd panel and the left half of the 3rd panel from the
    top, respectively.}
  \label{fig:17}       
\end{figure*}

\subsection{Galactic Plane survey}

\hess\ has also performed a survey of the Galactic Plane within a band of $-3 <
b < 3$~degrees in Galactic latitude $b$ (see Fig.~\ref{fig:17}) with a 3 hr
exposure per field. The scan resulted in about 40 sources of at least $5
\sigma$ significance (status 2007) in a band $280 < l < 60$~degrees of Galactic
longitude $l$, most of them spatially extended \cite{galscan07}. About half
of these sources are unidentified in other wavelength ranges. One gets the
impression that the amazing number of sources and variety of source types
almost fills the nearby part of the plane, as the significance map
Fig.~\ref{fig:17} suggests. The source types range from Pulsar Wind Nebulae,
Pulsars in binary systems, X-ray Binaries, shell-type SNRs and OB associations
to giant molecular clouds and the Galactic Center region. The survey is
presently continued. The number of VHE sources, at least in this special part
of the sky, appears mainly limited by the sensitivity of the instrument and
possibly also by confusion, rather than by the availability of sources.


\begin{figure*}[htb]
  \includegraphics[width=\textwidth]{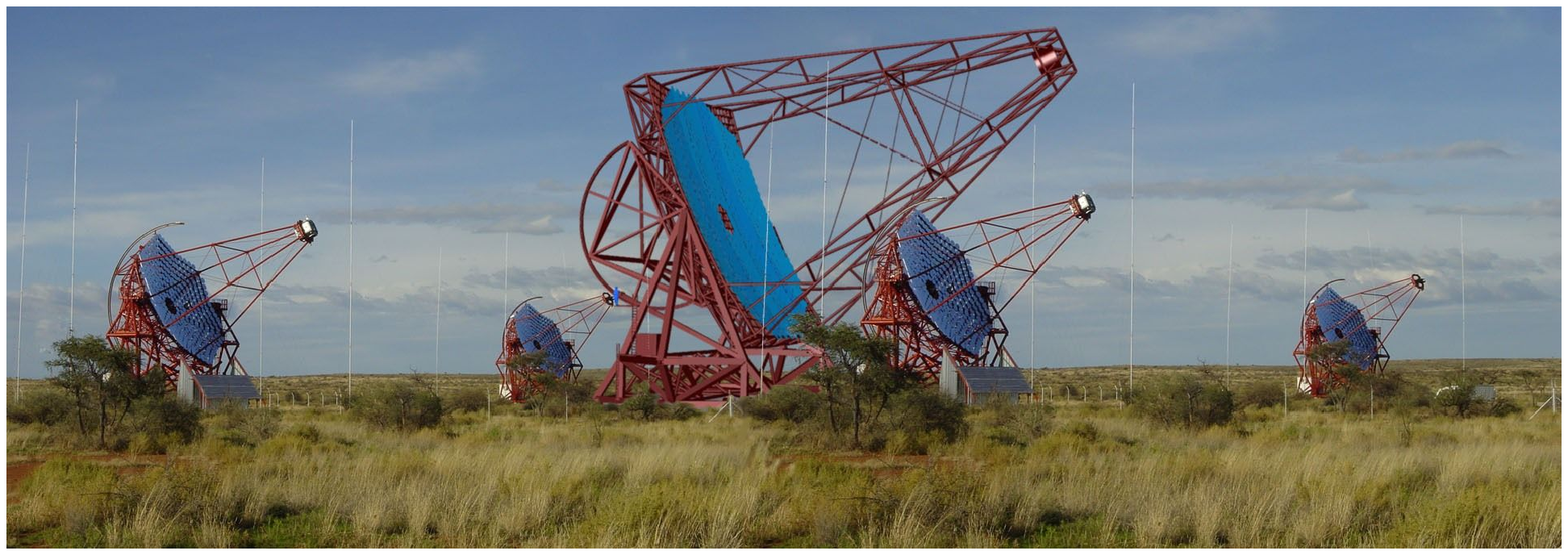}
  \caption[Photomontage of \hess\ Phase II]{Photomontage of \hess\ Phase II (courtesy W.~Hofmann). 
    The 28~m telescope in the center of
    the original \hess\ array should increase the sensitivity of the system by
    a factor $\approx 2$, at a threshold energy of $\sim 70$~GeV in coincidence
    mode.}
  \label{fig:18}       
\end{figure*}

\begin{figure*}[htb]\sidecaption
  \includegraphics[width=0.7\textwidth]{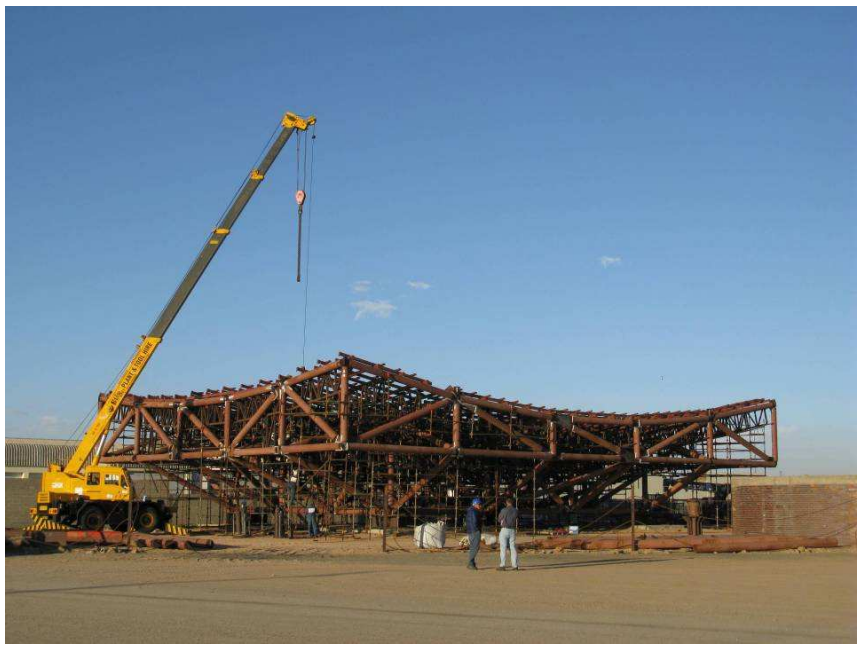}
  \caption[\hess\ II dish]{Build-up, in Namibia, of the dish and mirror support structure for
    the \hess\ II telescope (courtesy M.~Panter). 
    First light is expected at the end of 2009.}
  \label{fig:19}       
\end{figure*}

\section{The future}

The fact that the VHE sky is rich in sources - also in extragalactic space, as
recent detections by several observatories have shown - has led to efforts to
increase the power of the existing detectors. MAGIC started to experiment with
advanced solid state detectors in telescope cameras and to build a second
telescope, in order to be able to do stereoscopic observations. This second
telescope is currently - fall 2008 - in the commissioning phase (see
Fig.~\ref{fig:11}). \hess\ began in 2006 to build a 28~m telescope in the
center of the original \hess\ array (see Fig.~\ref{fig:18}), in order to lower
the threshold of the expanded stereoscopic system and to prototype and test the
capabilities of a significantly larger telescope (see Fig.~\ref{fig:19}).

Beyond these extensions of existing instruments, there are worldwide activities
to build a fourth generation of arrays. In the European context, with Japanese
participation, joint plans for a large array are in an advanced stage. The
project is called Cherenkov Telescope Array (CTA) \cite{cta} which should
enhance the sensitivity at 1 TeV by a further order of magnitude, down to the
level of $10^{-3}$ of the flux from the Crab Nebula (1 mCrab). In the US the
project Advanced Gamma-ray Imaging System (AGIS) \cite{agis} is actively
pursued, with a similar goal.

\begin{figure*}
\begin{center}
  \includegraphics[width=0.85\textwidth]{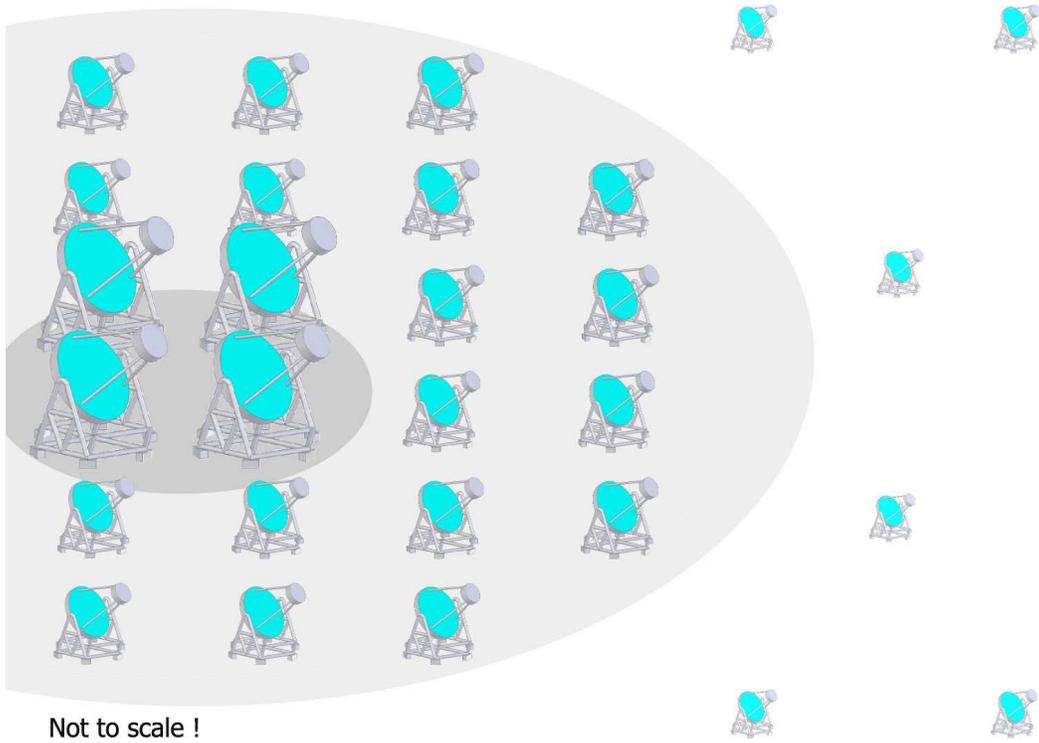}
\end{center}
  \caption{Possible configuration of the CTA project, involving a mix of
   telescopes of various sizes and characteristics (courtesy W.~Hofmann).}
  \label{fig:20}       
\end{figure*}

The CTA project involves different types and sizes of telescopes (see
Fig.~\ref{fig:20}) designed to extend the present energy range in both
directions: to lower energies of some tens of GeV and to higher energies in
excess of 100 TeV. Apart from deep broadband investigations of Galactic
objects, this should also allow the investigation of the non-thermal aspects of
large-scale structure formation in the Universe by studying active galactic
nuclei, clusters of galaxies, and the related phenomena of accretion,
starbursts, and galaxy-galaxy mergers. These goals clearly require the coverage
of a large energy range.

\begin{acknowledgements}
  We are indebted to our colleagues in the HEGRA and \hess\ Collaborations, and
  especially in the Max-Planck-Institut f\"ur Kernphysik for many discussions
  over the years on the physics of \gr\ observations with imaging Cherenkov
  telescopes. We also thank W.~Hofmann, G.~Hermann and M.~Panter for allowing
  us to use illustrative figures which they had designed for other purposes.
\end{acknowledgements}

\clearpage

\end{document}